\documentclass[amssymb,aps,twocolumn,showpacs,nofootinbib,superscriptaddress]{revtex4}
\usepackage[]{graphicx}
\usepackage[]{amsmath}
\usepackage{hyperref}
\usepackage{color}

\def\ket#1{| #1 \rangle}

\def\kb#1#2{| #1 \rangle\!\langle #2 |}

\def\cC{\mathcal{C}}
\def\cE{\mathcal{E}}

\def\cL{\mathcal{L}}

\def\cP{\mathcal{P}}

\def\cS{\mathcal{S}}

\def\fs{\mathfrak{s}}
\def\fl{\mathfrak{l}}
\def\ft{\mathfrak{t}}

\DeclareMathOperator*{\argmax}{\rm{argmax}}
\def\Eq#1{Eq.~\eqref{eq:#1}}

\begin{document}

\title{On the iterative decoding of sparse quantum codes}
\author{David Poulin\footnote{Present address: D\'epartement de Physique, Universit\'e de Sherbrooke, QC, Canada.}}
\email{David.Poulin@USherbrooke.ca}
\affiliation{Center for the Physics of Information, California Institute of Technology, Pasadena, CA 91125 USA.}
\author{Yeojin Chung}
\email{ychung@mail.smu.edu}
\affiliation{Department of Mathematics, Southern Methodist University, Dallas, TX 75205 USA.}

\date{\today}

\begin{abstract}
We address the problem of decoding sparse quantum error correction codes. For Pauli channels, this task can be accomplished by a version of the belief propagation algorithm used for decoding sparse classical codes. Quantum codes pose two new challenges however. Firstly, their Tanner graph unavoidably contain small loops which typically undermines the performance of belief propagation.  Secondly, sparse quantum codes are by definition highly degenerate. The standard belief propagation algorithm does not exploit this feature, but rather it is impaired by it.  We propose heuristic methods to improve belief propagation decoding, specifically targeted at these two problems. 
While our results exhibit a clear improvement due to the proposed heuristic methods, they also indicate that the main source of errors in the quantum coding scheme remains in the decoding. 
\end{abstract}

\pacs{}

\maketitle

\section{Introduction}

Quantum information science studies the information transmission and processing capabilities of quantum mechanical systems. It is now well established that quantum mechanical devices can in principle outperform classical ones in these two settings. The physical realization of such schemes, however, relies on our ability to cope with imperfect noisy systems. The theory of quantum error correction (QEC) and its fault-tolerant implementation~\cite{Sho95a,Ste96c,BDSW96a,Got97a,KL97a,KLZ98a,Pre99a} shows that this challenge is surmountable. The vast majority of studies of QEC and fault-tolerance has focused on small size block codes encoding one or very few qubits, and concatenation thereof. Noticeable exceptions are topological codes \cite{Kit03a,DKLP02a} (that have a vanishing rate), convolutional codes \cite{OT03a,OT04a,FGG05a,GR06a}, low density parity check (LDPC) codes \cite{MMM04a,COT05a,COT05b,SS07a}, and turbo-codes \cite{PTO07a}.

The main difficulty with small block QEC codes is that in order to achieve high error suppression, each logical qubit needs to be encoded in a very large number of physical qubits (the encodings are of low rates). This overhead poses an extra technical challenge for the design of practical quantum information processors. In classical coding theory, it is well known that smaller overheads can be achieved by encoding multiple bits in a single large block. In fact, random codes on $n$ bits typically achieve the Shannon capacity in terms of maximum rate at vanishing error probability as $n$ tends to infinity. The drawback, however, is that in general these codes cannot be decoded efficiently, i.e., computing the optimal recovery given an error syndrome is an NP-complete problem.

Two approaches can be pursued to overcome the decoding problem. Algebraic codes (e.g. Reed-Solomon codes \cite{RS60a}, Bose-Chaudhuri-Hocquenghem codes \cite{Hoc59a,BC60a}) are designed to have large minimal distances and efficient minimal distance decoders. Their construction relies on the structure of finite-field algebra, and they come an infinite variety of rates and minimal distances. Algebraic codes have been investigated in the quantum setting, e.g. \cite{Ste99d,AP02a,KKKS06a,AKS07a}. As argued e.g. in \cite{CF06a} however, due to their strict focus on the minimal distance, these codes are bound to fall short of the Shannon limit. They are on the other hand unequaled for channels with burst errors, and have therefore found numerous commercial applications.

The other approach is probabilistic coding (e.g., low density parity check [LDPC] codes, turbo-codes) which differs from algebraic coding in three important aspects. Firstly, the codes are chosen at random from a specified ensemble, and one is interested in the typical behavior over the ensemble. Secondly, minimum distance decoding is typically a hard problem on these code ensembles. This leads to the third important distinction which is a shift of focus from minimal distance to the average performance of the code ensemble under (sub-optimal) polynomial-time decoding.  For some common classical channels, state-of-the-art probabilistic codes are near-capacity achieving. The randomness involved in the code design yields good codes while the structure imposed on the code ensemble is sufficient to ensure good decoding performances. Probabilistic codes have also been investigated in the quantum context \cite{Cha98a,OT04a,MMM04a,OT05b,COT05a,FGG05a,GR06a,PTO07a}.

Probabilistic codes are decoded by means of a belief propagation (BP) algorithm (also known as sum-product, message passing, Viterbi algorithm, Bahl-Cocke-Jenilek-Raviv algorithm depending on context). BP is a highly parallel general-purpose algorithm to solve (or approximate the solution to) inference problems involving large number of random variables located at the vertices of a graph, where the edges of the graph encode dependency between the random variables \cite{AM00a,Yed01a,Mac03a}. When the graph is a tree, BP produces the exact solution to the inference problem in a time bounded by the tree's depth. More importantly, BP has proven to be a good heuristic method in cases where it is not known to converge, and particularly when the typical size of the loops in the graph is large. Recently, a generalization of BP was proposed to solve inference problems involving a large number of {\em quantum} systems \cite{LP07a,Has07b}. It was also shown that decoding a stabilizer QEC code on any memoryless quantum channel reduces to a quantum inference problem on a (generally loopy) graph. 

Sparse codes (a.k.a. LDPC) are designed to accomplish a good performance under BP decoding. They are obtained by imposing random linear constraints on codewords, with each constraint involving a small number of bits and similarly each bit being involved in a small number of constraints. The constraints can be represented by a Tanner graph (c.f. Fig.~\ref{fig:Tanner}). As a consequence of the sparseness of the constraints, the graph typically contains no or very few small loops. Decoded with a BP algorithm, sparse codes are amongst the best known classical error correction codes. For this reason, it is desirable to generalize them to the quantum setting. 

A method due to Calderbank, Shor, and  Steane \cite{CS96a,Ste96b} (CSS) enables the construction of QEC codes from any pair of dual classical codes. This technique has been used extensively to design small block codes, and recently to design ``hybrid" (small block flip, large block phase) QEC codes tailored for particular channels \cite{IM06a}. One might expect the CSS construction to leverage the power of sparse codes to QEC, but attempts in that direction have proven quite difficult \cite{MMM04a,COT05a}.  The main obstacle in the design of sparse CSS codes is that sparse classical codes --- being good error-correcting codes --- do not typically have sparse duals. Thus, randomly generated sparse codes are inappropriate and more sophisticated generating techniques are required for their use in CSS constructions.

The duality condition --- and, more generally, the commutation conditions imposed on any stabilizer codes, not just those obtained from the CSS construction --- also poses two immediate complications for the decoding of sparse QEC codes. On one hand, it implies that the graph representing a QEC code must have weight-4 loops. This places BP decoding on a slippery slope. On the other hand, when a classical code has a sparse dual code, it has by definition low weight errors. Fortunately, errors that lie in the dual code act trivially on the encoded quantum information. Any pair of errors that are related by an element of the dual code cannot and need not be distinguished: they are both corrected by the same operation. These are called ``degenerate errors" and by definition the sparse quantum codes have many degenerate low-weight errors. The degeneracy is a feature that generally improves the performance of QEC codes \cite{SS96a,DSS98a,SS06a}. Unfortunately, BP does not exploit the degeneracy of the code and even worse, it is typically impaired by degeneracy.  

The primary purpose of this paper is to describe why degeneracy compromises BP decoding. Then, in an attempt to alleviate this problem, we propose heuristic techniques to partially overcome the aforementioned obtacles. Our proposed techniques yields  significant improvements over standard BP decoding for the cases considered here. We note, however, that despite these improvements, our methods are in need of further development for achieving optimal performance in broad range of applications.

The rest of the paper is organized as follows. In the next section, we introduce the necessary background on QEC and set up some notation. Sec.~\ref{sec:BP} {provides} a detailed description of the BP algorithm as used for the decoding of sparse quantum codes. The following section describes the concept of degeneracy in QEC, and presents a simple model to explain why it should affect the performances of BP decoding. In Sec.~\ref{sec:techniques} we present some heuristic methods to improve BP decoding. Then, we present our numerical results in Sec.~\ref{sec:results}.

\section{Notation and background}
\label{Sec:stab}

\subsection{Stabilizer codes}

A QEC code on $n$ qubits is a subspace $\cC$ of the Hilbert space $(\mathbb{C}^2)^{\otimes n}$. It can be specified as the $+1$ common eigenspace of some set of commuting operators $S_1,S_2,\ldots S_m$ that generate under multiplication the so-called stabilizer group $\cS = \langle S_c \rangle$. Of particular interest are codes whose stabilizer generators are (up to a phase) $n$-fold tensor products of $2\times 2$ Pauli matrices
\begin{equation*}
I = \left(
\begin{array}{cc}
1&0 \\ 0 & 1
\end{array}
\right)
 X = \left(
\begin{array}{cc}
0&1 \\ 1 & 0
\end{array}
\right) 
Y = \left(
\begin{array}{cc}
0&-i \\ i & 0
\end{array}
\right)
Z = \left(
\begin{array}{cc}
1&0 \\ 0 & -1
\end{array}
\right),
\end{equation*}
and we will refer to such operators simply as Pauli operators. They form a group $\cP_n$ called the Pauli group. The elements of $\cP_n$ either commute or anti-commute, and for all $E,F \in \cP_n$, we define 
\begin{equation}
E\cdot F =
\left\{
\begin{array}{ll}
1 & {\rm if} \ \ EF = FE \\
-1 & {\rm if}\ \ EF = -FE.
\end{array}
\right.
\label{eq:comm}
\end{equation}
For $E,F \in \cP_1$, this operation is summarized in the following table.

\begin{center}
\begin{tabular}{|c|cccc|} 
\hline
$E\backslash F$ & $I$&$X$&$Y$&$Z$\\
\hline
$I$&1&1&1&1\\
$X$&1&1&-1&-1\\
$Y$&1&-1&1&-1\\
$Z$&1&-1&-1&1\\
\hline
\end{tabular} 
\end{center}
For $n$-qubit Pauli operators $E = E_1\otimes E_2\otimes \ldots \otimes E_n$ and $F = F_1\otimes F_2\otimes \ldots\otimes F_n$, we have
\begin{equation}
E\cdot F = \prod_{k=1}^n E_k\cdot F_k.
\end{equation} 

When $\cS$ has $m = n-k$ independent generators, $\cC \simeq (\mathbb{C}^2)^{\otimes k}$ will encode $k$ logical qubits. Operators that commute with the stabilizers elements have the property of mapping the code space to itself, and are therefore called logical operators. The logical operators also form a group $N(\cS)$, the normalizer of $\cS$ in $\cP_n$, which contains $\cS$ as a normal subgroup. The logical group is defined as $\cL = N(\cS)/\cS$ and coincides with $\cP_k$. 

For notational convenience, we will henceforth omit the tensor product symbol when describing an element of $\cP_n$. For instance, the element $X\otimes I\otimes I \otimes Y \in \cP_4$ will simply be denoted by $XIIY$. Alternatively, we use subscripts to identify the qubits that are acted upon by non-trivial Pauli matrices. For the above example, we would write $X_1Y_4$. 

The Clifford group on $n$ qubits $N(\cP_n)$ is composed of unitary matrices that map Pauli operators to Pauli operators --- it is the normalizer of $\cP_n$ in $U(2^n)$. One may succinctly describe a QEC code by a Clifford matrix $U$ acting on $n$ qubits as follows. The stabilizer generators are $S_c = UZ_c U^\dagger$ for $c=1,\ldots,m$. These operators clearly commute and are multiplicatively independent. From $U$, one can define the logical operators $\overline X_j = UX_{n-k+j}U^\dagger$ and $\overline Z_j = UZ_{n-k+j}U^\dagger$ for $j=1,\ldots k$. These operators commute with elements of $\cS$ and obey $\overline X_j\cdot \overline Z_{j'} = (-1)^{\delta_{j,j'}}$ and $\cL \simeq \langle \overline X_j,\overline Z_j\rangle$. With a slight abuse of notation, we will henceforth use $\cL$ to denote $N(\cS)/\cS$ and its representation given above. The code is then the set of all states of the form $U(\ket 0 ^{\otimes m} \otimes \ket\phi)$ where $\ket\phi \in (\mathbb{C}^2)^{\otimes k}$. It is also convenient to define a set of ``pure errors" by $T_c = UX_cU^\dagger$ for $c = 1,\ldots,m$. These operators commute with each other and with the logical operators, and obey the commutations $S_c\cdot T_{c'} = (-1)^{\delta_{c,c'}}$. The set $\{S_c,T_c,\overline X_j, \overline Z_j\}$ is a canonical generating set for $\cP_n$. 

\subsection{Tanner graph}

\begin{figure}[!tbh]
\includegraphics[width = 7cm]{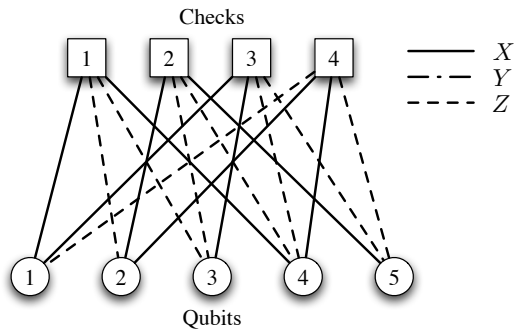}
\caption{Decorated Tanner graph for the 5-qubit code \cite{LMPZ96a,BDSW96a} with stabilizer generators $\{XZZXI$, $IXZZX$, $XIXZZ$, $ZXIXZ\}$. }
\label{fig:Tanner}
\end{figure}

A QEC code can be represented by a decorated Tanner graph. This is a bipartite graph $G = (V,E)$ with vertices $V = Q \cup C$ where the subset of vertices $Q$ represent the $n$ qubits, and the other subset $C$ represent the $m$ stabilizer generators, or checks. The graph has an edge $(q,c) \in E$ if and only if check $c$ acts non-trivially on qubit $q$, i.e. iff the $q$'th component of $S_c$ differs from the identity. Edge $(q,c) \in E$ is decorated by the $q$th Pauli matrix of check $c$, see Figure \ref{fig:Tanner}. When checks $c$ and $c'$ both act non-trivially on at least two qubits in common, say $q$ and $q'$, $G$ will contain a 4-loop $(c,q,c',q')$. 

To avoid the presence of 4-loops, one could make sure that no pair of checks $c$ and $c'$ act on more than one common qubits. The commutation condition between checks then implies that on that common qubit, $c$ and $c'$ must have the same Pauli matrix. This argument leads to the conclusion that every edge connected to qubit $q$ must carry the same decoration, say $Z$ for concreteness. Such a code, however, fails to detect the weight-1 error $Z_q$. This would not cause any problem if $Z_q \in \cS$, but in this case, qubit $q$ factors out in the code space, and the code states are of the form $\ket{\mu}\otimes\ket {0}_q$ for some $\ket{\mu} \in (\mathbb{C}^2)^{n-1}$ that satisfy additional constraints. We thus arrive at the conclusion that the Tanner graphs of QEC codes must unavoidably contain 4-loops. 

\subsection{Sparse quantum codes}

An edge in the Tanner graph is said to have qubit-degree $i$, if it is connected to a qubit node of degree (weight) $i$. Similarly, it has check-degree $i$, if it is connected to a check node of degree $i$. A sparse quantum code has a Tanner graph with qubit- and check-degree bounded by some constant independent of the number of encoded qubits. Since the checks must obey the strict commutation relations, it has proven difficult to generate them pseudo-randomly as it is done for the classical sparse codes. 

More precisely, a degree distribution $\gamma(x) = \sum_i\gamma_i x^{i-1}$ is a polynomial with nonnegative real coefficients satisfying $\gamma(1)=1$. Here, $\gamma_i$ denotes the fraction of edges in a graph for nodes of degree $i$. Let $\lambda_i$ and $\rho_i$ denote the fraction of edges of qubit-degree and check-degree $i$, respectively. Then, the degree distribution for the Tanner graph is the pair $(\lambda, \rho)$, where $\lambda = \sum_{i\geq 1} \lambda_{i}x^{i-1}$ and $\rho = \sum_{i \geq 1} \rho_i x^{i-1}$. The degree distribution $(\lambda,\rho)$ defines a code ensemble of rate $R = 1- \int_0^1 \rho(x)dx / \int_0^1 \lambda(x)dx$. We note that, in the classical theory of LDPC codes, a typical code in that ensemble decoded with the simple BP decoder on the binary erasure channel with erasure probability $\delta$ shows successful performance if $\delta \lambda (1-\rho(1-x)) < x$, for $x \in (0, \delta)$ \cite{LMS+97a,LMSS98b,RU01c}.

\section{Belief propagation decoding}
\label{sec:BP}

In this article, the error model that we consider are Pauli channels that have the form
\begin{equation}
\cE(\rho) = \sum_{E \in \cP_n} p(E) E \rho E^\dagger,
\label{eq:channel}
\end{equation}
where $p(E) \geq 0$ and $\sum_E p(E) = 1$. It is clear from Eq.~\eqref{eq:channel} that the phase of Pauli operators are irrelevant, so the sum should only be carried over the effective Pauli group $\cP_n/\{\pm 1, \pm i\}$. But to keep the notation simple, we will henceforth implicitly assume that $\cP_n$ is quotiented by its center.  A memoryless Pauli channel is one for which the probability factors as  $p(E) = \prod_{q=1}^n p_q(E_q)$ for all $E = E_1E_2\ldots E_n \in \cP_n$. A particularly relevant example is the depolarizing channel for which $p_q(I) = 1-\epsilon$ and $p_q(X) = p_q(Y) = P_q(Z) = \epsilon/3$ for all $q$, and for some depolarizing strength $0\leq \epsilon\leq 1$.

Once the $n$ qubits are prepared in a code state $\ket\psi \in \cC$, they are sent through the channel and state $\rho = \cE(\kb\psi\psi)$ is obtained by the receiver. To detect the possible errors at the receiver, the stabilizer generators $S_c$ are measured and outcome $s = (s_1,s_2,\ldots, s_m) \in \{-1,1\}^m$, called the error syndrome, is obtained.  The state $\rho$ is a statistical mixture of states $E\ket\psi$ for different $E \in \cP_n$. When the error $E$ that corrupted the register commutes with $S_c$, the syndrome bit $s_c$ takes value +1 because $S_c E \ket\psi = ES_c\ket\psi = E\ket\psi$. When $E$ anti-commutes with $S_c$ on the other hand, we obtain $S_cE\ket\psi = -ES_c\ket\psi = -E\ket\psi$, and hence, $s_c = -1$. Thus, the error syndrome provides partial information about the error that has occurred during the transmission. 

 Of all the errors $E \in \cP_n$ that could have corrupted the $n$ qubits, only those that obey a commutation relation with the generators of $\cS$ consistent with the error syndrome are to be considered after the measurements of the $S_c$'s. In other words, the probability distribution of the errors conditioned on error syndrome $s$ is 
\begin{equation}
p(E|s) \propto p(E) \prod_{c=1}^m \delta_{s_c,E\cdot S_c}.
\end{equation}
One decoding technique consists in identifying the most likely error given the syndrome $s$:
\begin{equation}
E^*(s) = \argmax_{E \in \cP_n} \{p(E|s)\}.
\label{eq:decoder}
\end{equation}
Since all elements of $\cP_n$ square to the identity (up to an irrelevant phase), the information can then be corrected by applying $E^*(s)$ to the register. We will see in the next section however that this is not the optimal decoding rule. 

The computationally difficult step in the above error correction scheme is the evaluation of $E^*(s)$. In fact, this problem is  NP-complete \cite{BMT78a}.\footnote{This problem is actually equivalent to that of finding the ground state of a spin-glass, which gives rise to a host of connections between sparse code theory and statistical physics \cite{MM07a}.} Thus, we will settle for a simpler task that consists of evaluating the qubit-wise most-likely error. For $E = E_1\cdots E_n$, let us define the marginal probabilities 
\begin{equation}
p_q(E_q|s) = \sum_{E_1,\ldots  E_{q-1},E_{q+1},\ldots E_n} p(E_1\ldots E_n|s).
\label{eq:marginal}
\end{equation}
The qubit-wise most likely error is then 
\begin{equation}
E_q^*(s) = \argmax_{E_q \in \cP_1} \{p_q(E_q|s)\}.
\end{equation}
In general, the marginal optimum $E_1^*(s) \ldots E_n^*(s)$ needs not coincide with the global optimum $E^*(s)$. 

Evaluating $E^*_j(s)$ is also a difficult task because Eq.~\eqref{eq:marginal} involves a sum over a number of terms exponential in $n$.  This is where belief propagation comes in handy.\footnote{As noted above, Pauli channels, being defined in terms of a probability over $\cP_n$, make use of a classical BP algorithm. For arbitrary quantum channels, a quantum BP algorithm was described in details in \cite{LP07a}. }  This algorithm operates by sending messages along the edges of the Tanner graph (c.f. Fig.~\ref{fig:Tanner}). Messages from qubit $q$ to check $c$ are denoted by $m_{q\rightarrow c}$ and messages from check $c$ to qubit $q$ are denoted $m_{c\rightarrow q}$. Messages received at and sent by qubit $q$ are  probability distributions over $E_q$. In other words, they are arrays of 4 positive numbers, one for each value $E_q = I, X, Y,$ and $Z$. We denote $n(q)$ the neighbors of qubit $q$, and define $n(c)$ similarly. 

To initialize the algorithm, each qubit sends out a message to its neighbors equal to its prior probability of error $m_{q\rightarrow c}(E_q) = p_q(E_q)$, where $p_q(E_q)$ is the probability entering in the definition of the memoryless channel \Eq{channel}. Upon reception of these messages, each check sends out a message to its neighboring qubits given by
\begin{equation}
m_{c \rightarrow q}(E_q) \propto \!\!\!\sum_{\substack{E_{q'} \\ q' \in n(c)\backslash q} } \!\!\!\!\!\!\Big( \delta_{s_c, S_c\cdot E_c} \prod_{q' \in n(c)\backslash q} \!\!\! m_{q'\rightarrow c}(E_{q'}) \Big),
\label{eq:m_cq}
\end{equation}
where $n(c)\backslash q$ denotes all neighbors of $c$ except $q$. The sum is over all Pauli operators on the neighbors of $c$ but with $E_q$ held fixed. The proportionality factor can be fixed by normalization $\sum_{E_e} m_{c\rightarrow q} = 1$. Thus, $m_{c \rightarrow q}$ is a function of the syndrome bit $s_c$ associated to check $c$, and the messages $m_{q'\rightarrow c}$ received from all neighbors of $c$, except $q$. 

Upon reception of these messages, each qubit sends out a message to its neighboring checks given by
\begin{equation}
m_{q\rightarrow c}(E_q) \propto p(E_q)\!\!\prod_{c' \in n(q) \backslash c}\!\! m_{c'\rightarrow q}(E_q),
\label{eq:m_qc}
\end{equation}
where $n(q)\backslash c$ denotes all neighbors of $q$ except $c$. Again, the proportionality factor can be fixed by normalization. Thus, $m_{q\rightarrow c}$ is a function of the qubit prior probability $p_q(E_q)$, and the messages $m_{c'\rightarrow q}$ received from all neighbors of $q$, except $c$. 

Equations (\ref{eq:m_cq}-\ref{eq:m_qc}) define an iterative procedure that is at the core of BP. The beliefs $b_q(E_q)$ --- that are meant to represent an approximation to the marginal conditional probability $p_q(E_q|s)$ \Eq{marginal} ---  are computed from the $m_{c\rightarrow q}$ messages as follows
\begin{equation}
b_q(E_q) = p_q(E_q) \prod_{c \in n(q)} m_{c\rightarrow q}(E_q). 
\label{eq:beliefs}
\end{equation}
The recovery can be chosen as the product of qubit-wise maximum-belief Pauli matrices $E^{BP} = \bigotimes_{q=1}^n E^{BP}_q$ where
\begin{equation}
E_q^{BP} = \argmax_{E_q \in \cP_1} \{b_q(E_q)\}.
\end{equation}

When $G$ is a tree, the beliefs converge to the correct conditional marginals $b_q(E_q) = p_q(E_q|s)$ in a time equal to the tree's depth. Thus, the qubit-wise most likely error can be computed efficiently. This leads, for instance, to an efficient maximum-likelihood decoding algorithm for concatenated quantum block codes \cite{Pou06b}. 

When the graph $G$ includes loops, the beliefs will not converge to the correct conditional marginals in general. Nonetheless BP can be used as a heuristic method to approximate the solution to such intrinsically hard problems, and in many situations it is empirically observed to provide reliable solutions.  Apropos examples of successful loopy BP are provided by the near-capacity achieving decoding of classical LDPC and turbo-codes. 

Since loopy BP is not guaranteed to converge within any fixed number of iterations, one has to impose a halting criterion on the iterative procedure. Here, we choose to halt when the correction given by the highest beliefs $E^{BP}$ lead to a trivial error syndrome, i.e., when $E^{BP}\cdot S_j = s_j$ for all $j=1,\ldots,m$. Since this is not guaranteed to occur after any finite number of iterations, it is also necessary to impose a maximum number of iterations.  

\section{Degeneracy}
\label{Sec:degeneracy}

As explained above, the output of the channel $\rho$ is a statistical mixture of states $E\ket\psi$ for different $E \in \cP_n$. Note however that when $E' = ES$ for some operator $S \in \cS$, then $E\ket\psi = E'\ket\psi$ by definition of code states. Moreover, $E$ and $E'$ have the same error syndrome since they differ by an element of $\cS$, which by definition commute with all the generators $S_c$. Hence, errors $E$ and $E'$ cannot be distinguished.  This is not a concern however because any recovery $R$ that reverses the effect of $E$ on the code space will also reverse the effect of $E'$
\begin{equation*}
RE\ket\psi = \ket\psi \Rightarrow RE'\ket\psi = RES\ket\psi = RE\ket\psi = \ket\psi.
\end{equation*}
In brief, the errors that are related by an element of $\cS$ {\em  cannot} and {\em need not} be distinguished by the error syndrome. This is the main feature of degeneracy in QEC. 

Since $\{S_c,T_c,\overline X_j, \overline Z_j\}$ is a canonical generating set for $\cP_n$, we can uniquely expand each error as $E = \fs(E)\ft(E)\fl(E)$ where $\fs(E) \in \cS$, $\ft(E) \in \langle T_c \rangle$, and $\fl(E) \in \langle \overline X_j, \overline Z_j \rangle$. Note that $\ft(E)$ is the only element of that decomposition that can anti-commute with the stabilizer generators. Moreover every elements of $\langle T_c\rangle$ has a distinct commutation relation with the $S_c$, and hence a distinct syndrome. It follows that $\ft(E)$ is entirely known given the error syndrome $s$, and it is given by 
\begin{equation}
\ft(s) = \prod_{c=1}^m T_c^{\frac{1+s_c}{2}}.
\end{equation}
On the other hand, the value of $\fs(E)$ is of no interest since that element of the decomposition acts trivially on $\cC$. The only element left is $\fl(E)$, and it has conditional probability
\begin{eqnarray}
p(L|s) &\propto& \sum_{E: \fl(E) = L} P(E) \\
&=& \sum_{S \in \cS} p(E = S\ft(s)L).
\label{eq:marginalL}
\end{eqnarray}
Optimal decoding consists in identifying the most likely value for $L$ given $s$
\begin{equation}
L^*(s) = \argmax_{L \in \cL} \{p(L|s)\}.
\label{eq:ML}
\end{equation}
This differs from the decoder defined at Eq.~\eqref{eq:decoder} because $p(L|s)$ is obtained from $p(E|s)$ by quotienting $\cP_n$ by $\cS$. In this sense, QEC codes are intrinsically coset codes.\footnote{In statistical physics terms, decoder Eq.~\eqref{eq:decoder} minimizes the energy while Eq.~\eqref{eq:ML} minimizes the free energy.}

The degeneracy typically improves the performance of a code because many errors can be corrected in the same way, and need not be distinguished by the error syndrome. This becomes particularly relevant when many elements of $\cS$ have low weight since the degeneracy will then relate typical errors. To take advantage of this feature, it is important to properly sum the probabilities over the cosets: the coset with the most likely error may differ from the most likely coset. However, the BP algorithm presented in the previous section uses Eq.~\eqref{eq:decoder} as a starting point, and so completely ignores the coset structure of the code. This will lead to a decline of the scheme's performances, and should be particularly pronounced for sparse codes.

Beyond these complications, the degeneracy is expected to cause extra complications when used with any qubit-wise maximum-likelihood decoder such as  BP. This is due to the presence of symmetries in the code that are reflected in the decoder, but broken by the channel. The next section aims at illustrating this issue. 

\subsection{A case study}
\label{sec:2qubits}

We consider a particularly simple QEC code that encodes zero qubits (a single state) into two qubits. It has stabilizer generators $XX$ and $ZZ$, and code space $\cC = \{(\ket{00} + \ket{11})/\sqrt 2 \}$. Suppose these qubits are sent through a depolarization channel, and that error $IX$ is applied. Measurement of the stabilizer generators reveal the error syndrome $s = (+1,-1)$. This error can be corrected by either of the operations $XI$, $IX$, $YZ$, or $ZY$. Because of the symmetry of the code however, the marginal conditional probabilities Eq.~\eqref{eq:marginal} are identical on both qubits, $P_1(E|s) = P_2(E|s)$. Since none of the correct recoveries have this symmetry, any decoding scheme based on qubit-wise probabilities will fail with this very simple code.

\begin{figure*}[!tbh]
\includegraphics[width = 17cm]{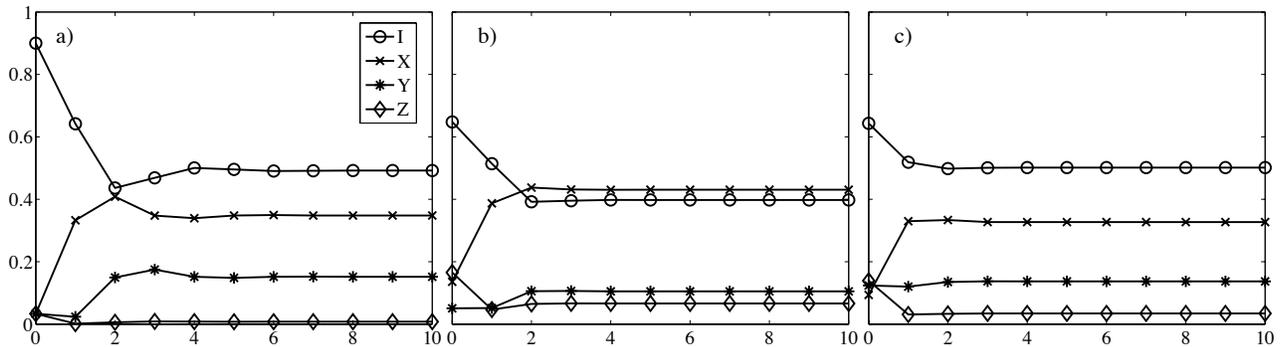}
\caption{Beliefs as a function of number of iterations for a QEC code with stabilizer generators $XX$ and $ZZ$ with syndrome $(+1,-1)$. a) The prior $p_q(E_q)$ is the depolarizing channel with $\epsilon = 0.1$ for both qubits. In that case, $E^{BP} = II$ which yields a detected error. On b) and c), a random perturbation of strength $\delta = 1$ was applied to the prior,  breaking the symmetry between the two qubits. b) shows the beliefs for qubit 1 and c) the beliefs for qubit 2.  In this case, $E^{BP} = XI$ which is an appropriate recovery.}
\label{fig:loop}
\end{figure*}

Figure~\ref{fig:loop} a) shows the beliefs obtained from the BP algorithm as a function of the number of iterations. Due to the symmetry of the code, the beliefs on qubit 1 and 2 are identical. As we see, the identity is always assigned the largest belief, and hence, the BP decoder will recommend correcting the error by doing nothing. This is, of course, incorrect. In fact, this correction will result in a ``detected error" because there will remain a non-trivial syndrome after applying the recommended recovery. In that case, our BP algorithm would iterate until it reaches the maximum number of iterations imposed by hand, and would output the correction $II$ knowing that it is an incorrect inference. In fact, the decoder cannot possibly succeed because it always assigns identical beliefs to qubits 1 and 2, but none of the appropriate corrections listed above have that symmetry. 

The optimal decoder must add the probability of $XI$ and $IX$, and assign the result to either one of them; this choice is not important and can be made at random. This is not incorporated in the standard BP scheme because $XI$ and $IX$ are symmetric in all aspects. To achieve any improvement in the BP decoding, it is essential to break this symmetry. 

\section{Heuristic methods for degenerate codes}
\label{sec:techniques}

While the example of the previous section may appear naive and contrived, it actually captures a universal feature of sparse quantum codes: they have many degenerate typical errors, and the degenerate errors of equal weight are completely symmetric under BP decoding. One needs to break this symmetry in order to overcome the difficulty explained in the previous section. We have investigated a host of different techniques to improve the performance of sparse quantum codes under BP decoding. Here, we present the most successful ones. 

In all cases, the general structure of the BP algorithm is unaltered. At each round of iteration, new beliefs are obtained and the maximum-belief correction $E^{BP}$ is evaluated. If $E^{BP}$ restores a trivial syndrome, the algorithm halts with output $E^{BP}$. If, on the other hand, the procedures reaches a pre-determined fixed number of iterations $T_{pert}$ without satisfying the halting conditions, a perturbation will be applied in order to break any symmetry that may be responsible for the impasse. The different techniques we have explored differ in the choice of the symmetry breaking. The presence of small loops in the Tanner graph turns out to be crucial because it rapidly propagates the symmetry breaking perturbation to all the qubits involved in the degenerate error. 

\subsection{Freezing}
\label{sec:freeze}

The first technique consists of ``freezing" one of the qubits involved in an unsatisfied check. We first find a frustrated check $c$, i.e., $S_c \cdot E^{BP} \neq s_c$. We pick a random qubit $q$ connected to that check, and freeze its prior probability to $p_q(E_q) = \delta_{E_q,I}$. We let the BP iterate with this modified prior for $T_{pert}$ steps. If it halts during this period we are done. If $c$ is still frustrated, we restore the probability of $q$ and freeze a different qubit $q'$ involved in $c$. If $c$ is not frustrated but the halting condition is not satisfied, we find another frustrated check $c'$ and freeze one of its qubits at random.  

When applied to the above example, the random freezing technique immediately solves the problem. If for instance we set $p_2(E_2) = \delta_{E_2,I}$, we obtain after a single iteration $b_1(E_1) = \delta_{E_1,X}$ and $b_2(E_2) = \delta_{E_1,I}$, hence the correction $E^{BP} = XI$. 

\subsection{Random perturbation}
\label{sec:RP}

The second technique consists in randomly perturbing the prior probability associated to qubits involved in frustrated checks. We simply identify the frustrated checks $c$, and for each $q$ in $n(c)$ we apply the perturbation (up to normalization)
\begin{eqnarray}
p_q(I_q) &\rightarrow& p_q(I_q) \label{eq:pert1}\\
p_q(X_q) &\rightarrow& (1+\delta_X)p_q(X_q) \\
p_q(Y_q) &\rightarrow& (1+\delta_Y)p_q(Y_q) \\
p_q(Z_q) &\rightarrow& (1+\delta_Z)p_q(Z_q) \label{eq:pert2}
\end{eqnarray}
where $\delta_X,\delta_Y,$ and $\delta_Z$ are random variables distributed uniformly in the range $[0,\delta]$ for some fixed $\delta$. The goal of this perturbation is to create an asymmetry among the qubits that will put an end to the impasse. It is clear that the perturbation needs to be random in order to break the symmetry: if the error probability is equally increased on all unsatisfied checks the problem would persist. It also seems intuitive to move to a prior with strictly stronger error probability because at least one of the qubits connected to $c$ definitely has an error. This choice is also justified by the fact that for the vast majority of decoding errors, the decoder had identified the identity operator as the most likely error. In other words, we have empirically observed that the decoder is naturally too biased towards the identity.  

When applied to the toy model of the previous section, a random perturbation produced exactly the effect we were hoping for, as long as the perturbation creates a sufficiently strong asymmetry between $p_1(X)$ and $p_2(X)$. An example illustrates this point on Fig.~\ref{fig:loop} b)-c). 

\subsection{Collision}
\label{sec:collide}

The collision trick is an attempt to enforce the analogy between real decoding instances and the simple model presented in Sec. \ref{sec:2qubits}. It operates in conjunction with other symmetry breaking techniques. It proceeds by finding a colliding pair of unsatisfied checks $c$ and $c'$, such that both $ S_c\cdot E^{BP} \neq s_c$ and $S_{c'}\cdot E^{BP} \neq s_{c'}$, and $c$ and $c'$ share some qubits in common. Typically, $c$ and $c'$ will have two qubits in common. Under the  assumption that the non-trivial syndromes $s_c$ and $s_{c'}$ have a common cause --- i.e. that they are due to error on their common qubits --- the situation reduces exactly to that of Sec.~\ref{sec:2qubits}. Thus, we can apply a random perturbation to those qubits, or choose to freeze one of them. It is also possible to further exploit the labels of the checks on each qubits to determine the local degeneracy structure of the code. 

\section{Results}
\label{sec:results}

We illustrate the performances of our decoding algorithms with a bicycle sparse quantum code~\cite{MMM04a} used over a depolarizing channel. We have applied our decoding algorithms to a number of other codes; in all cases we have found that our heuristic methods yield an improvement over the standard BP decoder but the best performances were obtained from the bicycle codes. 

Bicycle codes~\cite{MMM04a} are CSS codes that make use of cyclic matrices. A simple method to create a $d \times d$ cyclic binary matrix ${\bf C}$ consists of generating a binary column vector ${\bf A}$ of length $d$ with random entries and defining $({\bf C})_{i,j} = ({\bf A})_{i+j}$. A sparse cyclic matrix is obtained when the vector ${\bf A}$ is sparse. 

For a bicycle code with row-weight $w$, block length $n$ and number of checks $m$, first generate a random $n/2 \times n/2$ cyclic matrix ${\bf C}$ with row weight $w/2$. (As in \cite{MMM04a}, this matrix can be constructed from a difference set, but the construction outline above seems to work just as well.) Then, construct a matrix ${\bf H}_0$ by merging the matrices ${\bf C}$ and ${\bf C}^\dagger$, i.e., ${\bf H}_0 = ({\bf C}|{\bf C}^{\dagger})$. This matrix is self-dual:
\begin{eqnarray}
({\bf H}_0{\bf H}^{\dagger}_0)_{i,j} &=& ({\bf C}{\bf C}^{\dagger} + {\bf C}^{\dagger}{\bf C})_{i,j}\\
&=& \sum_k {\bf A}_{i+k}{\bf A}_{j+k} + {\bf A}_{k+i}{\bf A}_{k+j} = 0.
\end{eqnarray}
Deleting $n/2-m/2$ of the rows from ${\bf  H}_0$ yields a self-dual matrix ${\bf H}$ with $m/2$ rows and $n$ columns. This matrix is used in the CSS construction to obtain a code of rate $\frac{n-m}{n}$ with checks given by
\begin{equation}
S_c = 
\left\{\begin{array}{ll} 
\prod_{j=1}^n Z^{({\bf H})_{c,j}} & \ {\rm for}\ 0 < c \leq m/2, \\
\prod_{j=1}^n X^{({\bf H})_{c,j}} & \ {\rm for}\ m/2 < c \leq m.
\end{array}\right.
\end{equation}
The commutativity of the checks follows from the self-duality of $\bf H$.

The construction of bicycle codes has complete freedom in controlling the size and weight of the code. However, all the deleted rows of ${\bf H}_0$ are low-weight codewords which are unlikely to lie in the dual. Hence, while they offer good performances under BP decoding on the depolarization channel, bicycle codes most likely have a small minimal distance (less or equal to $w$).

\begin{figure}[!tbh]
\includegraphics[width = 8cm]{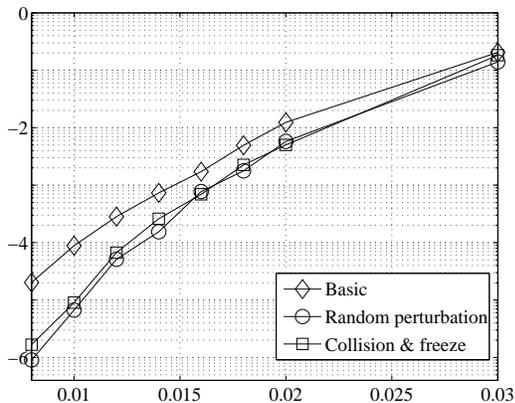}
\caption{Block error probability {\it vs} depolarization strength $\epsilon$ for three different decoding procedure as explained in the text. The code encodes 400 logical qubits in 800 physical qubits, for a rate of 0.5. The checks all have degree 30 and the average degree of the qubits is 15. The maximum number of iterations is 90, the number of iterations between each perturbations is $T_{pert} = 6$, and the strength of the random perturbation [c.f. Eqs.~(\ref{eq:pert1}-\ref{eq:pert2})] is $\delta = 0.1$. }
\label{fig:BC}
\end{figure}

The most successful methods for improving the iterative decoding performances are freezing (Sec.~\ref{sec:freeze}) combined with the method of colliding checks (Sec.~\ref{sec:collide}), and the simple random perturbation (Sec.~\ref{sec:RP}). Our results obtained for bicycle codes are shown at Fig.~\ref{fig:BC}. 
Both symmetry breaking techniques yield a 10 dB gain over the basic BP decoder at depolarizing strength 0.01 for the bicycle codes. In particular, the bicycle code has a decoding error below $10^{-4}$ around $\epsilon \approx 0.01$ using the basic BP decoder and this value is increased to $\epsilon \approx 0.014$ using our heuristic techniques  (c.f. Fig.~11 of Ref.~\cite{MMM04a}). Results obtained from larger blocks exhibit overall better performances, but the improvement due to the modified BP decoding are more pronounced on smaller blocks for the range of error probabilities accessible through Monte Carlo simulations. 

Note that there are two distinct types of failed error corrections, depending on whether $EE^{BP} \in N(\cS)$ or not. In the former case, the error is undetected because it corresponds to a codeword. These errors can generally be attributed to fundamental limitations of the code. In the latter case, it is known that the decoder has failed because the syndrome after the correction procedure remains non-trivial. Such a {\em detected error} indicates a failure of the decoder. Despite the improvements yielded by our heuristic methods, in all the cases we have investigated, 100$\%$ of the errors were detected errors. This indicates that the major obstacle to good quantum coding schemes remains in the decoding. The only exceptions were the codes presented in \cite{COT05a}, that we found to contain many weight-6 codewords.

\section{Conclusion}

In this paper, we have explained how the high degeneracy of sparse quantum codes will typically impair their performance under belief propagation decoding. Based on a simple model, we have proposed some heuristic techniques to partially overcome this problem. Our numerical results show that these techniques provide a clear and substantial improvement of the coding scheme's performance, which indirectly corroborates our model. 

Despite these efforts, we found that the greatest challenge still remains in the decoding. Indeed, all the errors found in our simulations could be attributed to the decoder rather than the finite minimal distance of the code. Thus, it is necessary to make further progress in the proposed decoding techniques for their broad application. 

\section{Acknowledgments}

We are grateful to H. Ollivier for valuable comments and useful discussions. DP is supported in part by the Gordon and Betty Moore Foundation through Caltech's Center for the Physics of Information, by the National Science Foundation under Grant No. PHY-0456720, and by the Natural Sciences and Engineering Research Council of Canada. YC is supported by University Research Council of Southern Methodist University.

%\bibliographystyle{/Users/dpoulin/archive/hsiam}
%\bibliography{/Users/dpoulin/archive/qubib}

\end{document}